\documentclass{JINST}
\usepackage{units}
\usepackage{textcomp} 
\usepackage{gensymb}
\usepackage{graphicx}
\usepackage{graphics}

\graphicspath{{plots/}}
\title{GEM Module Design for the ILD TPC}

\author{T. Behnke$^a$, F. M\"uller$^a$,
A. M\"unnich$^a$\thanks{Corresponding author.}~
and K. Zenker$^a$\\
\llap{$^a$}Deutsches Elektron Synchrotron DESY,\\
  D-22603 Hamburg, Germany\\
E-mail: \email{astrid.muennich@desy.de}}

\abstract{A Time Projection Chamber (TPC) using micro-pattern gas detectors is planned as the main tracking device for a detector at the next Linear Collider. A novel support structure for Gas Electron Multipliers (GEMs), which minimizes the material and improves the flatness of the foils, has been developed and tested with multiple GEM modules in a large TPC prototype at DESY. Reducing dead material at the GEM module boundaries improves the field homogeneity. In addition, it was shown in simulation that a field shaping ring at the border of the module can improve the charge collection in regions with non-homogeneous fields. This shaping wire was integrated into the module design and a successful test beam campaign with three modules has been carried out. First results regarding resolution and field distortions will be discussed.}


\begin{document}
\section{Introduction}\label{sec:intro}
An e$^+$e$^-$ Linear Collider has the potential for excellent precision measurements. Therefore the requirements for the detector performance are very high~\cite{DBD}. The detector concepts are optimized to utilize Particle Flow~\cite{PFA}. A central part of this strategy is a very good tracking system. The physics requirement for the momentum resolution is $\Delta\left(1/p_{\mathrm{T}}\right) = \unit[2\cdot10^{-5}]{GeV^{-1}}$.

\begin{figure}[t]
  \centering
  \includegraphics[width=\textwidth]{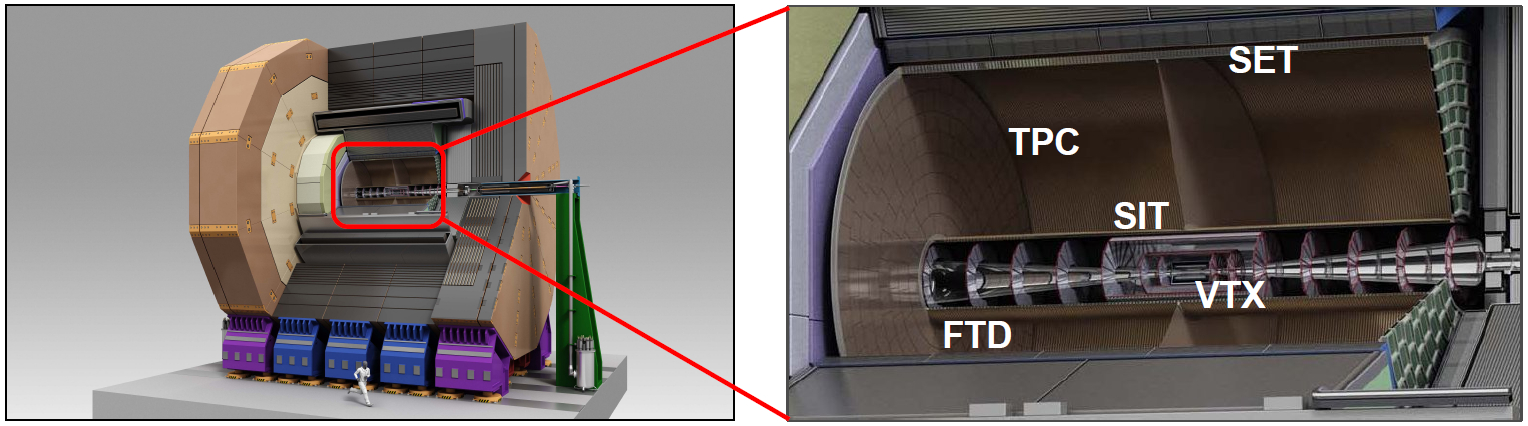}
  \caption{The ILD detector concept with its tracking region~\cite{DBD}. 
    The vertex tracker (VTX) is complemented by a Silicon Inner Tracker (SIT) 
    in the barrel region and Forward Tracking Detector (FTD) in the forward region. 
    The main tracker is the Time Projection Chamber (TPC) followed by a Silicon External Tracker (SET).}
  \label{fig:ILD}
\end{figure}

For the International Large Detector (ILD) a Time Projection Chamber (TPC) is planned as the central tracking device as depicted in Fig.~\ref{fig:ILD}. The TPC alone has to provide a momentum resolution of $\Delta\left(1/p_{\mathrm{T}}\right) = \unit[10^{-4}]{GeV^{-1}}$. This can be achieved with a point resolution of $\sim\unit[100]{\micro m}$ over the full drift length. With $\sim$ 200 position measurements along a particle track the TPC offers excellent pattern recognition capability and a tracking efficiency close to $\unit[100]{\%}$ down to low momenta.
In addition a very low material budget is placed in front of the highly segmented calorimeter especially in the barrel region where the TPC stays below $\unit[5]{\%}$ of a radiation length $X_0$.

In order to reach such a resolution micro-pattern gas detectors will be used for gas amplification at the anode. 
Gas Electron Multipliers (GEMs)~\cite{GEM} are one option and are used in this context to build several readout modules for a large TPC prototype.

\section{GEM Module Design}\label{sec:GEMModule}
\begin{figure}[t]
\centering
\includegraphics[width=.5\textwidth]{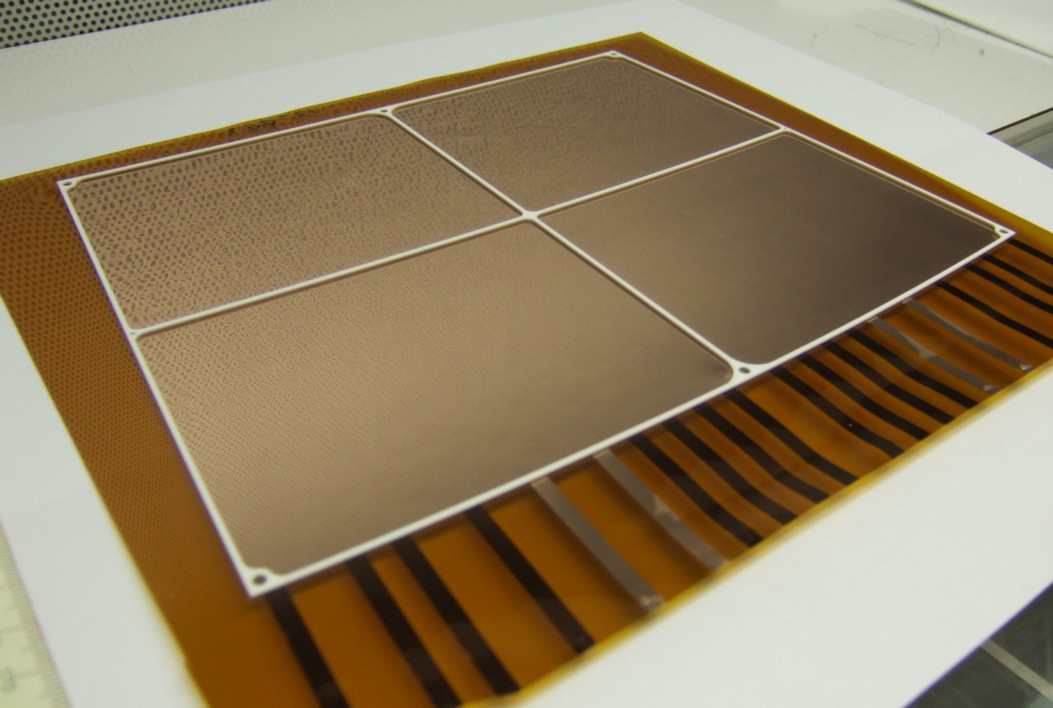}
\includegraphics[width=.4\textwidth]{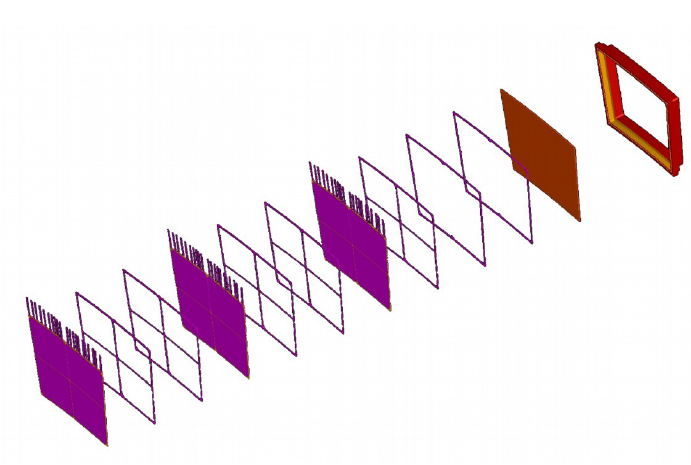}
\caption{On the left a GEM glued to its ceramic frame is shown. The exploded view of the triple GEM stack on the right illustrates that the frames serve as spacer between the GEMs as well.}
\label{fig:GEM}
\end{figure}

The main design goals of a GEM readout module are to limit the dead space and the material budget. Therefore a thin frame is necessary also implying that the flatness of the GEM has to be reached without stretching it due to the limited force the frame can tolerate~\cite{gemgrid}.
This thin ceramic frame is shown in Fig.~\ref{fig:GEM}.
In order to reach a reliable high voltage stability each GEM is divided into 4 sectors on the anode side. To limit field distortions between sectors the cathode side has no division. A triple GEM stack is chosen which allows for stable and versatile operation at high gain.

Based on a previous version for a GEM module~\cite{oldmodule} an improved design has been implemented.
The new generation of modules now have a fully sensitive read out plane with 4892 pads of a size of $\unit[1.26\times5.85]{mm^2}$.
Improvements to the high voltage distribution were done,
resulting in a very stable operation with only one spark in the very beginning of a two week measurement campaign.
In addition a shaping wire was attached to the grid frame of the upper GEM on the stack. With this addition potential of the wire the field distortions at the border of the modules can be reduced.

\section{The Large Prototype TPC}\label{sec:LP}
The Large Prototype (LP) has been built within the LCTPC collaboration~\cite{lctpc_coll,lctpc} to compare
different readout modules under identical 
conditions and to address integration issues. 
The main goals are to study the point and momentum resolution as well as $dE/dx$.

\subsection*{Field Cage}\label{sec:FC}
The field cage has a length of $\unit[61]{cm}$ and a diameter of $\unit[72]{cm}$. It can reach a cathode voltage of 
up to \unit[24]{kV} which for common gases corresponds to drift fields up to $\unit[350]{V/cm}$.
Composite materials were used to achieve a low material budget of $\unit[1.24]{\%}$ of a radiation length X$_0$~\cite{fieldcage}.

The endplate can hold up to seven modules each with a size of about $\unit[22\times17]{cm^2}$. The design of the endplate and the module placement resembles a cut out off a large scale endplate. A picture of the endplate can be seen in Fig.~\ref{fig:TB} on the left showing three of the slots equipped with GEM modules and the remaining four with termination shield modules.


\subsection*{Test Beam Setup}\label{sec:TB}
\begin{figure}[t]
\centering
\includegraphics[width=.4\textwidth]{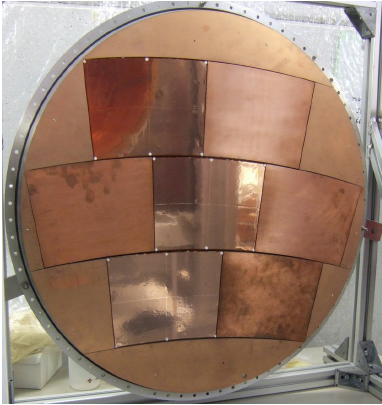}
\includegraphics[width=.345\textwidth]{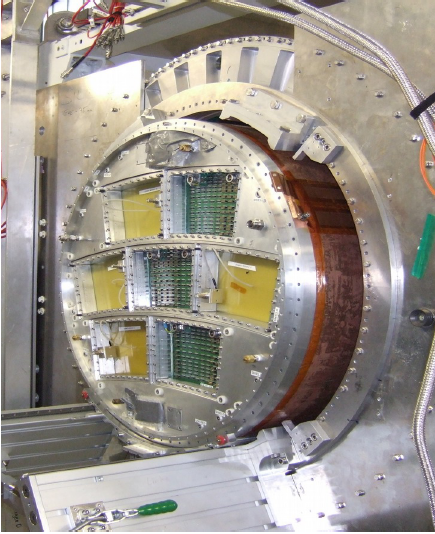}
\includegraphics[width=.11\textwidth]{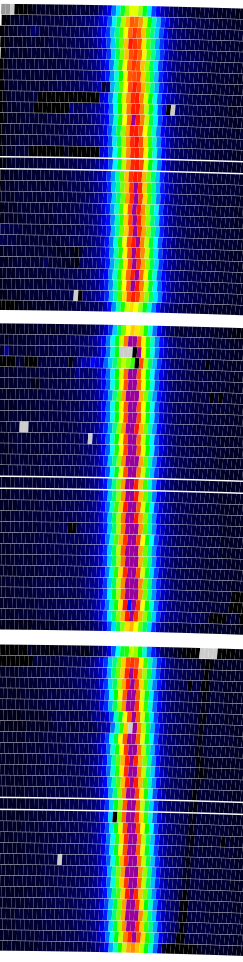}
\caption{Three GEM modules were installed in the endplate visible in the left picture as the shiny surfaces. The middle picture shows the LP inserted into the magnet before the electronics is attached. On the right an integrated beam profile is shown crossing the three modules.}
\label{fig:TB}
\end{figure}

The DESY test beam facility~\cite{TB_DESY} provides an e$^\pm$ beam up to $\unit[6]{GeV}$.
A $\unit[1]{T}$ magnet with a bore large enough to fit the LP is available.
The magnet is mounted on a movable stage, enabling a movement of the LP around three axes.
A test beam campaign was carried out with three modules of the new improved design introduced in the previous chapter.
Although the full module area is equipped with readout pads only one half of each module is equipped with electronics due to limited amount of available channels and space constraints. Pads which are not connected to a readout channel are grounded. In total 7200 channels were read out with the ALTRO electronics~\cite{ALTRO} with a shaping time of $\unit[120]{ns}$. A lever arm of about $\unit[50]{cm}$ along the beam was achieved with this configuration as shown in Fig.~\ref{fig:TB}.

\section{Field Simulations}\label{sec:Sim}
An electrostatic field simulation was carried out with CST\texttrademark~\cite{CST} to study the field behavior at the border of the GEM module. 
An interface from CST\texttrademark~\cite{CST_interface} to \textsc{Garfield++}~\cite{garfield} has been developed which is used to simulate the drift electrons in the electric field calculated by CST\texttrademark.

\begin{figure}[t]
\centering
\def\svgwidth{0.7\textwidth}
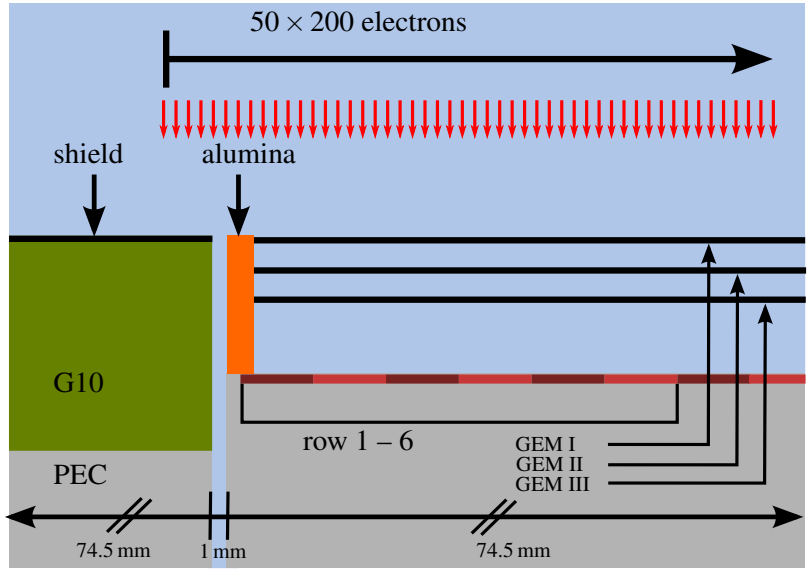
\caption{Simulation setup of the triple GEM stack and its border with the endplate. Electrons are started above the GEM to study the deviation of their drift path caused by the field inhomogeneities.}
\label{fig:SimSetup}
\end{figure}

Fig.~\ref{fig:SimSetup} shows the setup of the simulation which studies the behavior at the border between a GEM module and the endplate.

\begin{figure}[t]
\centering
\includegraphics[width=\textwidth]{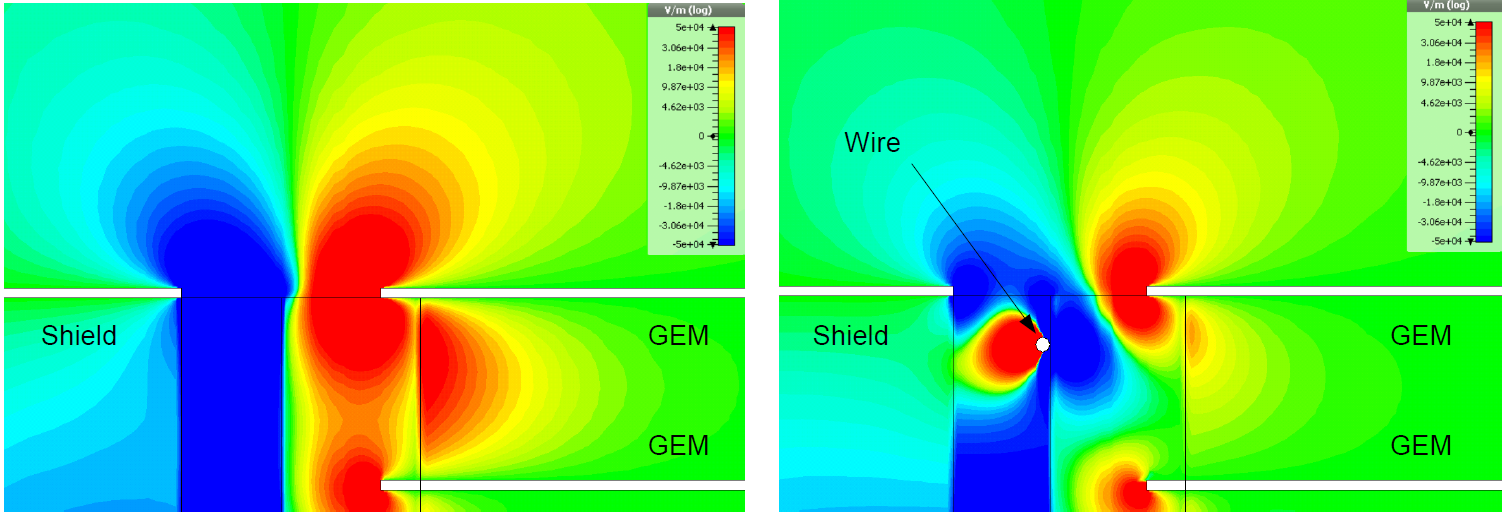}
\caption{Field simulation of the triple GEM stack with (right) and without (left) an additional field shaping wire. The electric field component transverse to the drift field is shown. The field inhomogeneity at the border of the module causes a displacement of primary electrons.}
\label{fig:SimField}
\end{figure}

\begin{figure}[t]
\centering
\def\svgwidth{0.7\textwidth}
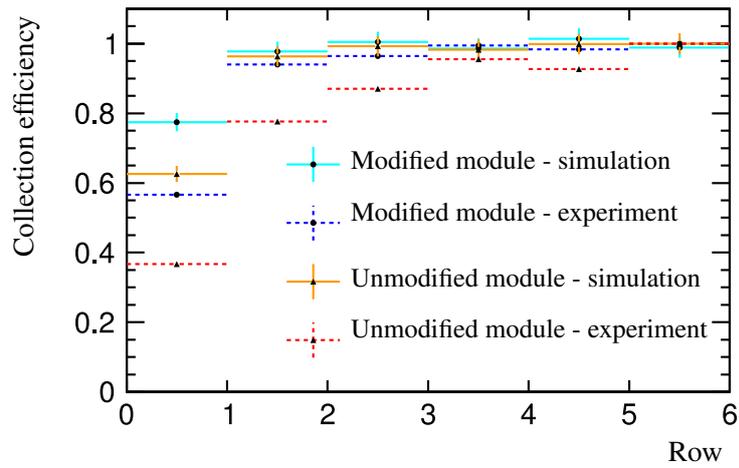
\caption{Comparison of the charge collection efficiency at $\unit[0]{T}$ with and without the addition of a field shaping wire and between simulation and measurement.}
\label{fig:SimEff}
\end{figure}

The simulations show that the addition of a field shaping wire reduces the field distortions shown in Fig.~\ref{fig:SimField}. The potential of the wire is set to cathode potential of the top GEM in the stack. Results for other configurations are presented in Ref.~\cite{CSTSim}. The charge collection efficiency at the border of the readout module could be improved by $\unit[20]{\%}$ both in simulation as well as in the measurement as shown in Fig.~\ref{fig:SimEff}. The relative agreement between simulation and measurement is quiet good when comparing the different module designs. The absolute values are different due to simplifications in the simulation (see Ref.~\cite{CSTSim}).

To study the distortion of the electron path, 200 electrons are started at 50 different positions along the pad rows but at the same distance above the module.

\begin{figure}[t]
\centering
\def\svgwidth{0.48\textwidth}
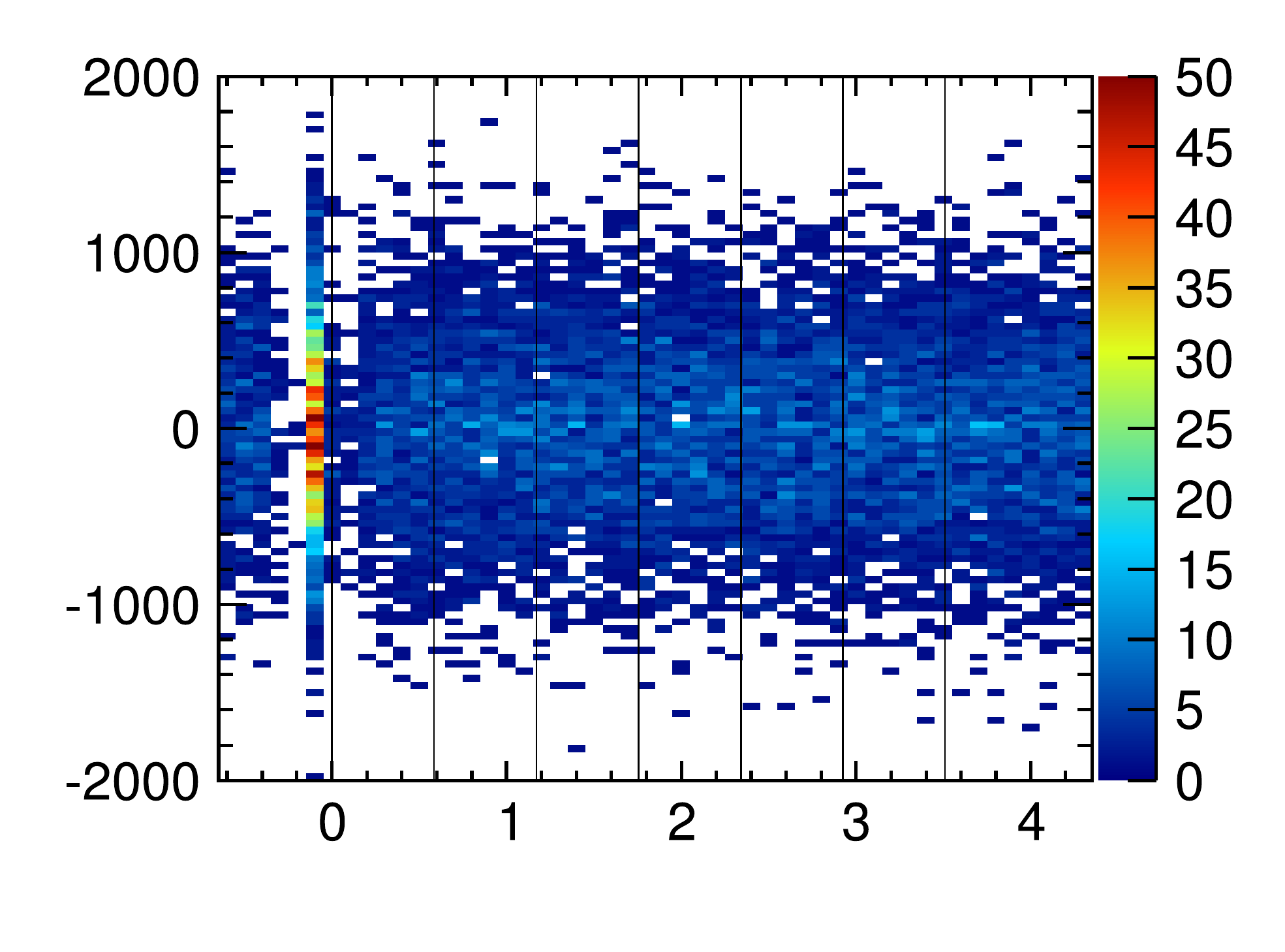
\hfill
\def\svgwidth{0.48\textwidth}
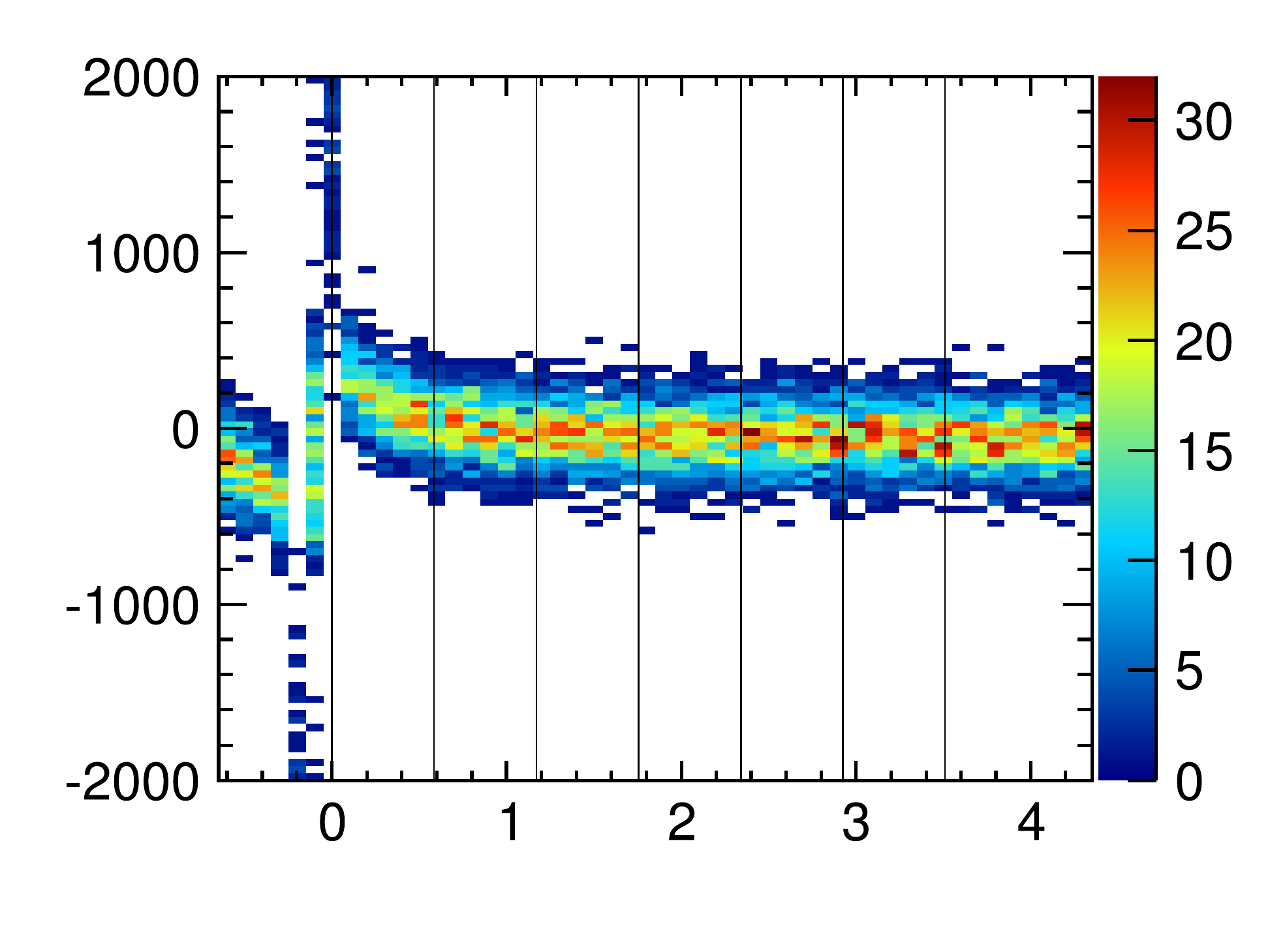
\def\svgwidth{0.48\textwidth}
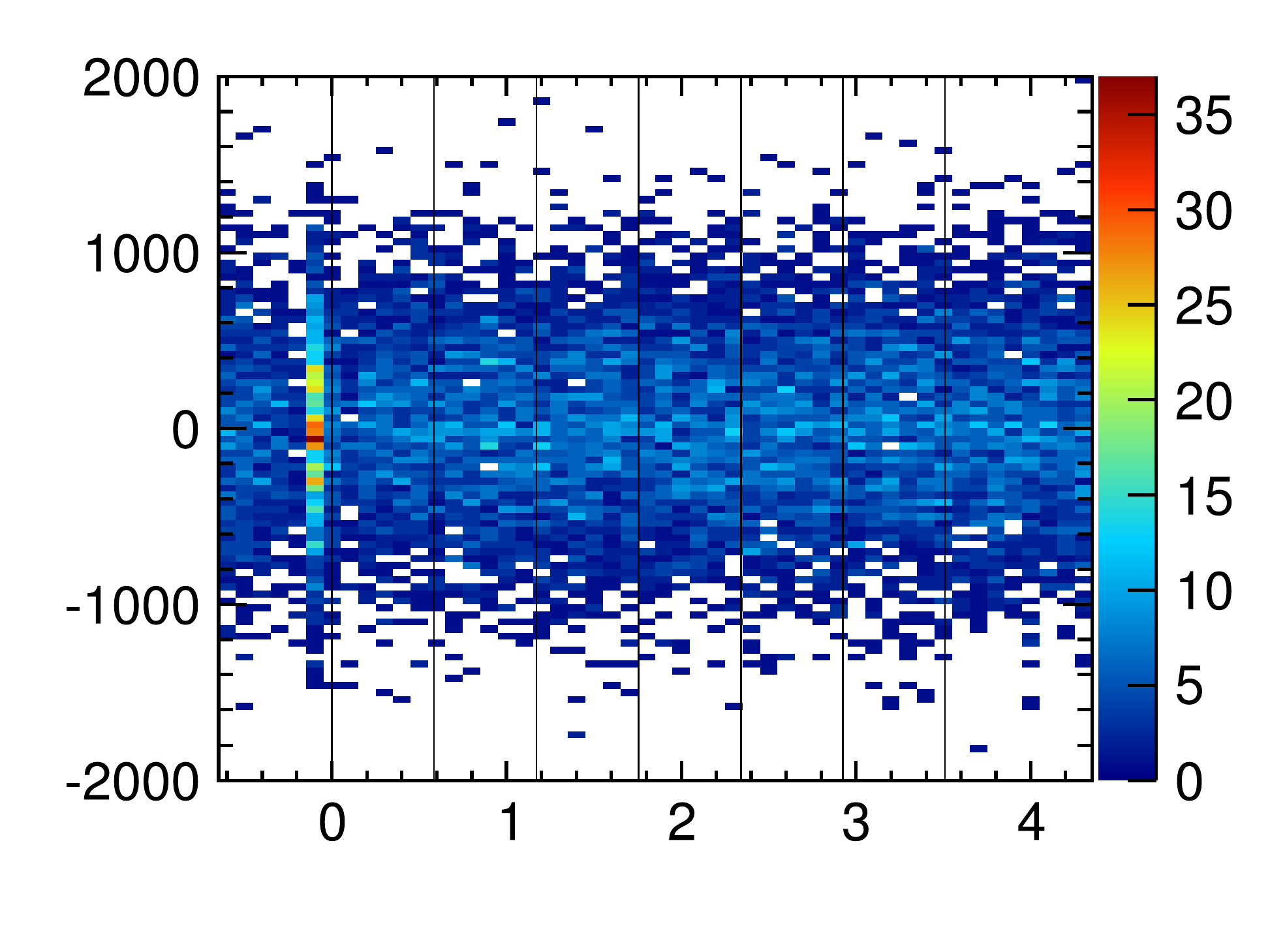
\hfill
\def\svgwidth{0.48\textwidth}
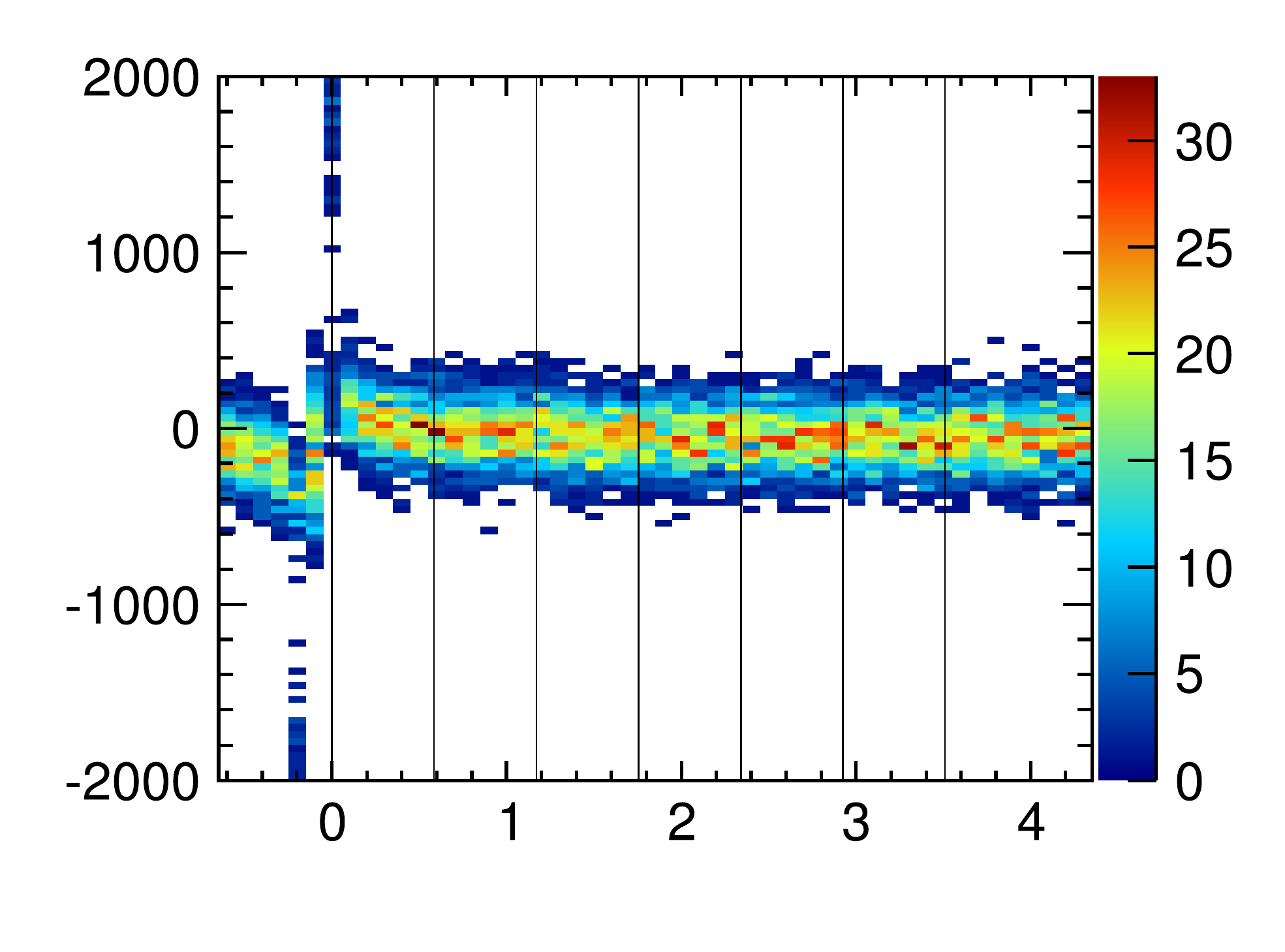
\caption{Deviation of the electron drift path due to field distortion for $\unit[0]{T}$ (left) and $\unit[1]{T}$ (right) magnetic field and for a GEM module without (top) and with (bottom) a field shaping wire. Black lines indicate readout rows of the module.}
\label{fig:SimResult}
\end{figure}

Without magnetic field the electrons follow the electric field lines and end up in the gap between modules. With the introduction of a magnetic field $E\times B$ effects change the electron drift path. This results in an increase of the charge collection efficiency but also in a distortion of the electron arrival positions near the module border. This behavior can be seen in Fig.~\ref{fig:SimResult}. The amplitude of the distortion is reduced with the shaping wire but still visible.


\section{Measurements}\label{sec:Meas}

\subsection*{Module Boundaries}\label{sec:dist}
\begin{figure}[t]
\centering
\includegraphics[width=0.7\textwidth]{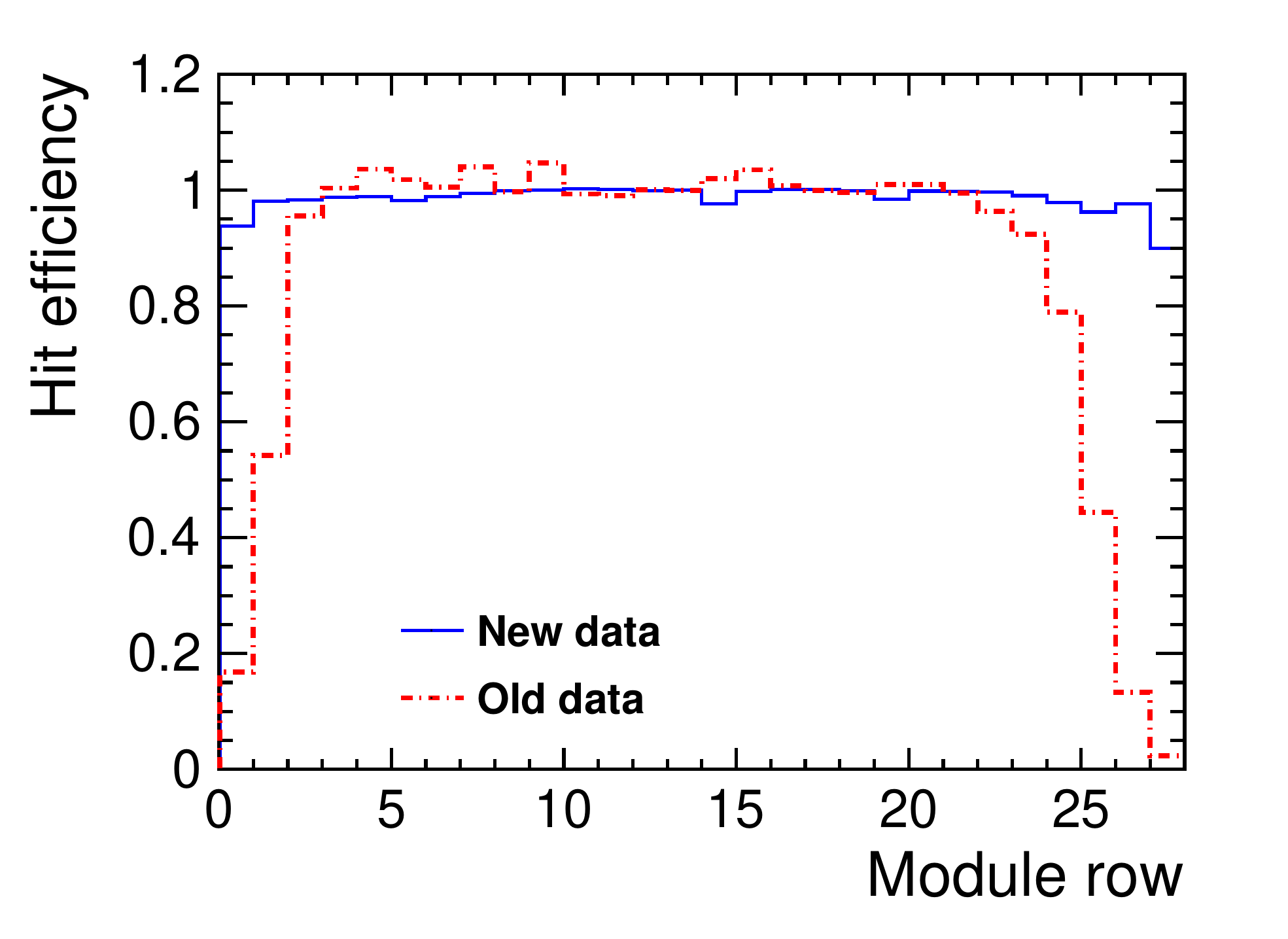}
\caption{Comparing the hit efficiency at $\unit[0]{T}$ for the old and new module design a significant improvement due to the shaping wire can be seen.}
\label{fig:hit_comp}
\end{figure}

The hit efficiency was improved to above $\unit[95]{\%}$ due to the guard ring which is a major improvement compared with the previous module design as show in Fig.~\ref{fig:hit_comp}. Each distribution was normalized to row 14 to allow for a comparison of the shape.
The charge efficiency is still lower at the module borders as could be seen in Fig.~\ref{fig:SimEff}.

\begin{figure}[t]
\centering
\includegraphics[width=0.7\textwidth]{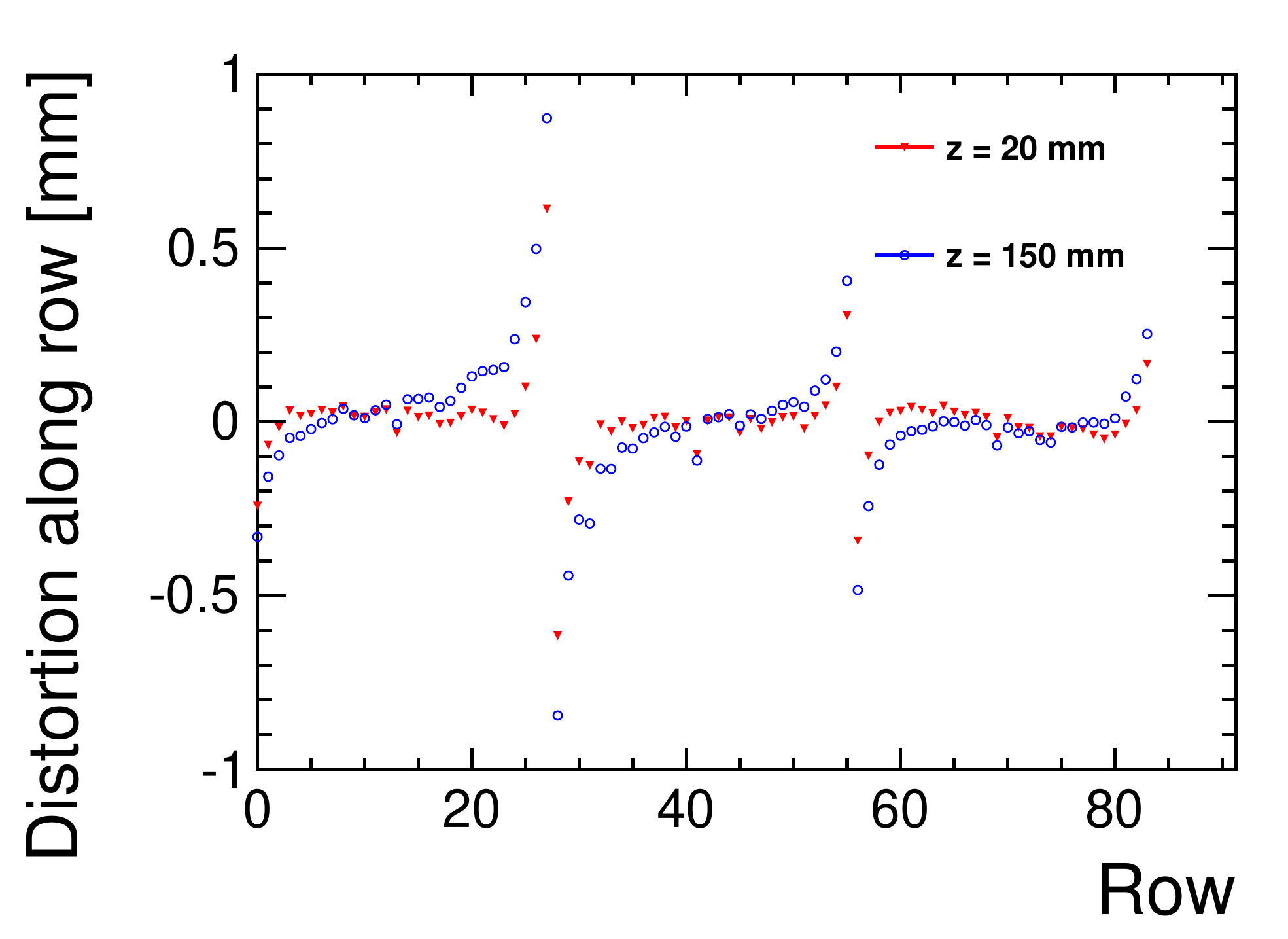}
\caption{The mean value of the distance from a reconstructed point to the track is shown in dependence on the row number. Each module has 28 rows. The distortions are shown for two different drift distances at $\unit[1]{T}$.}
\label{fig:dist}
\end{figure}

The field distortions at the boundaries of the modules are observed in the data. They lead to charge loss on the outer rows and bending of the drift path of the electrons due to $E\times B$ effects. Fig.~\ref{fig:dist} shows the deviation of the measured hit from the track for each row on each module for two different drift distances. The same behavior as in the simulation is observed but on a larger scale. The main contribution to the distortion stems from the module boundaries. A small dependence on the drift distance is visible. This can point to inhomogeneities during the drift process in the field cage. Further studies are needed to understand and quantify the dependencies and systematic effects contributing to the observed distortions. Without an external reference track the method itself is prone to be sensitive to the track fit and other possible reconstruction effects.

\subsection*{Single Point Resolution}\label{sec:res}
\begin{figure}[t]
\centering
\includegraphics[width=0.7\textwidth]{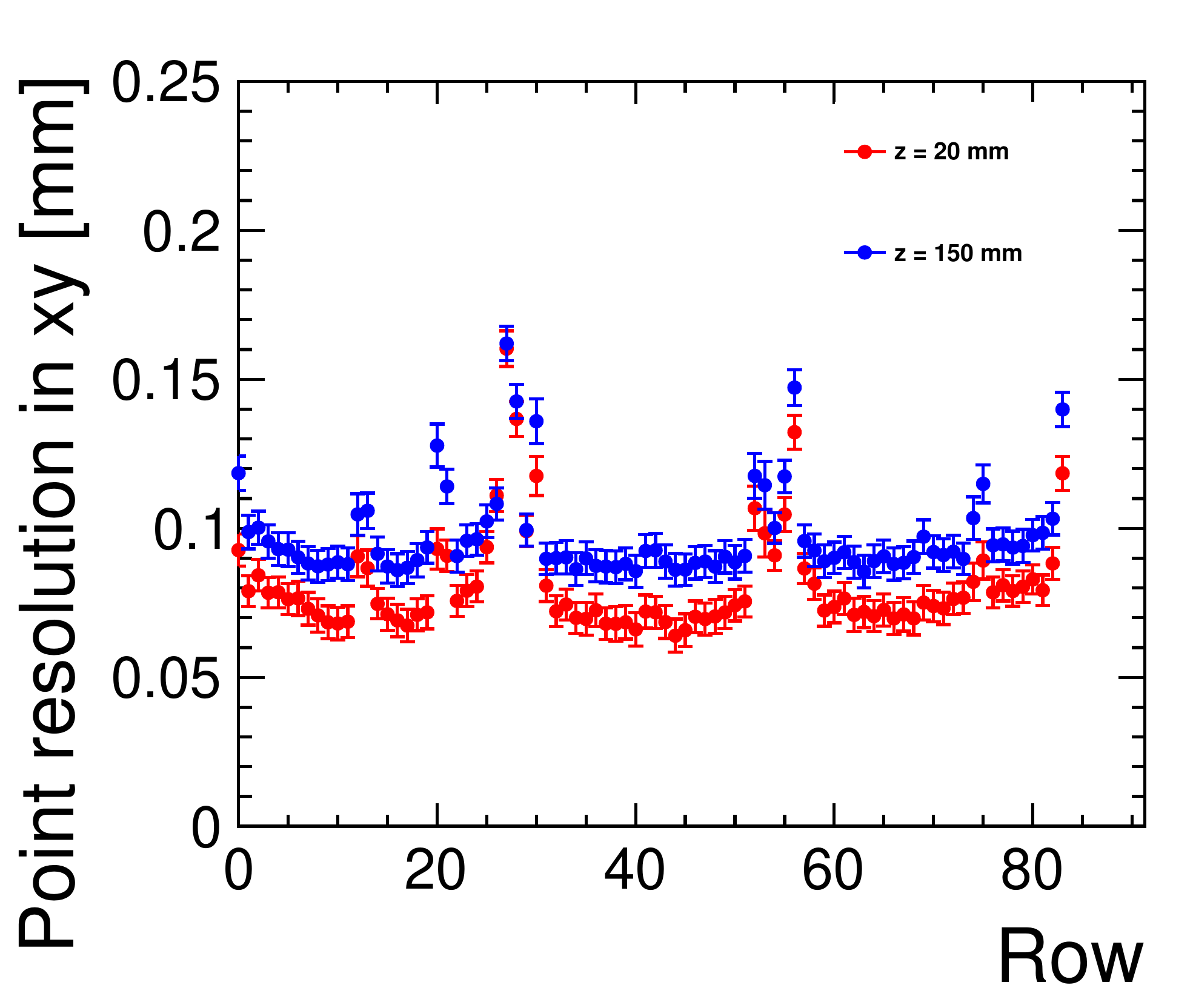}
\caption{Single point resolution in xy shown in dependence on the row number. Each module has 28 rows. The resolution is evaluated for two different drift distances at $\unit[1]{T}$.}
\label{fig:res_row}
\end{figure}

The single point resolution is sensitive to the observed field distortions. In Fig.~\ref{fig:res_row} the resolution is evaluated for each row on the three modules separately. The resolution is degraded in regions of large distortions as can be seen from a direct comparison with Fig.~\ref{fig:dist} showing the distortions for the same two data sets. This is most likely an effect of a change in track angle along the row due to the distortions. The data set contains tracks mainly perpendicular to the measurement rows. If the angle along a row increases the charge distribution is wider similar to the effect of diffusion. Therefore the resolution is less precise.

If one follows the standard approach and extracts the resolution from the width of a single distribution including all rows, this distribution will be broadened as it is an accumulation of Gaussian distribution with different mean values due to the distortions. To obtain a correct measurement of the resolution the distortions have to be corrected. At the moment only a data driven correction of the distortion evaluated on a run by run basis can be carried out.

\begin{figure}[t]
\centering
\includegraphics[width=0.48\textwidth]{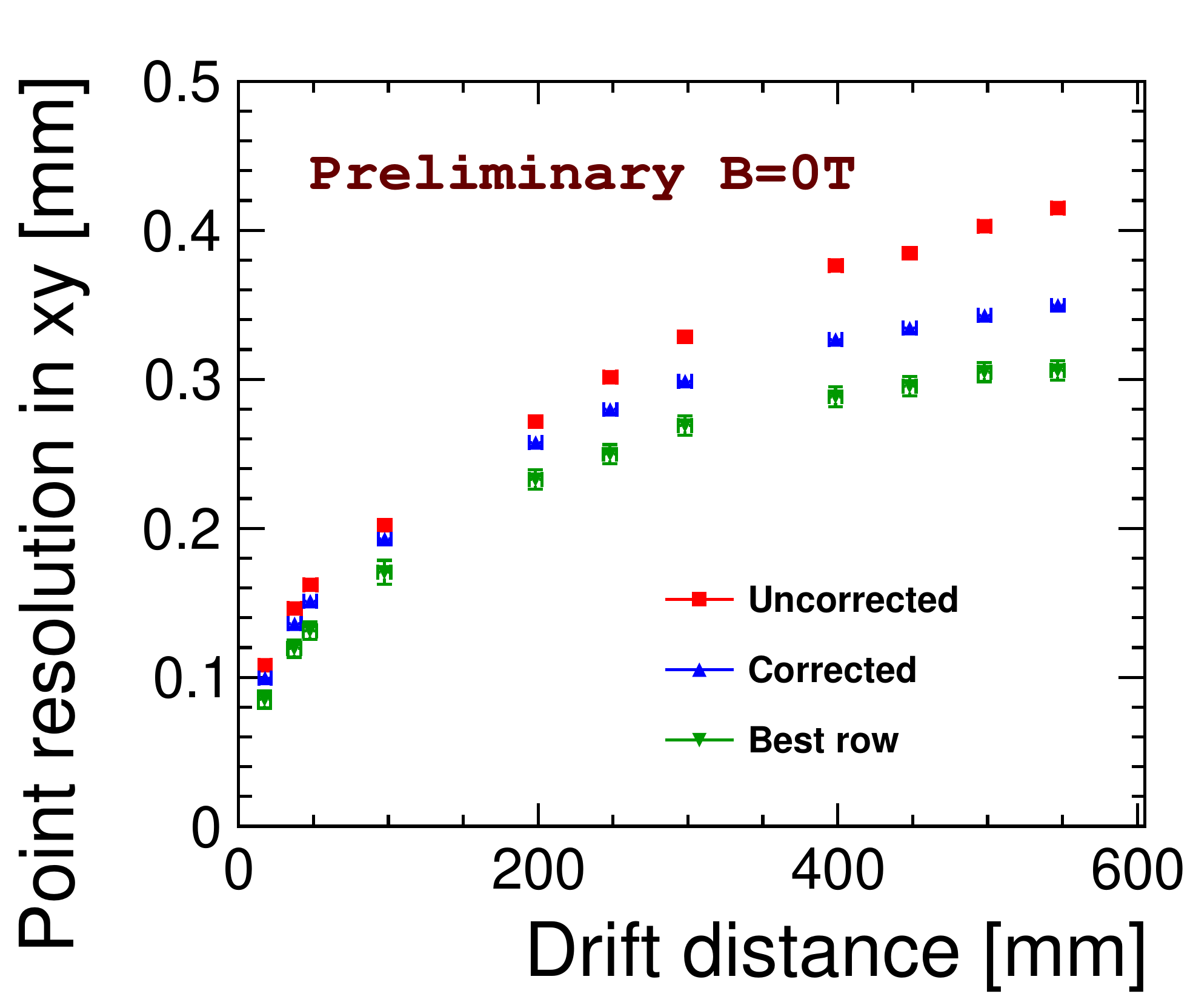}
\includegraphics[width=0.48\textwidth]{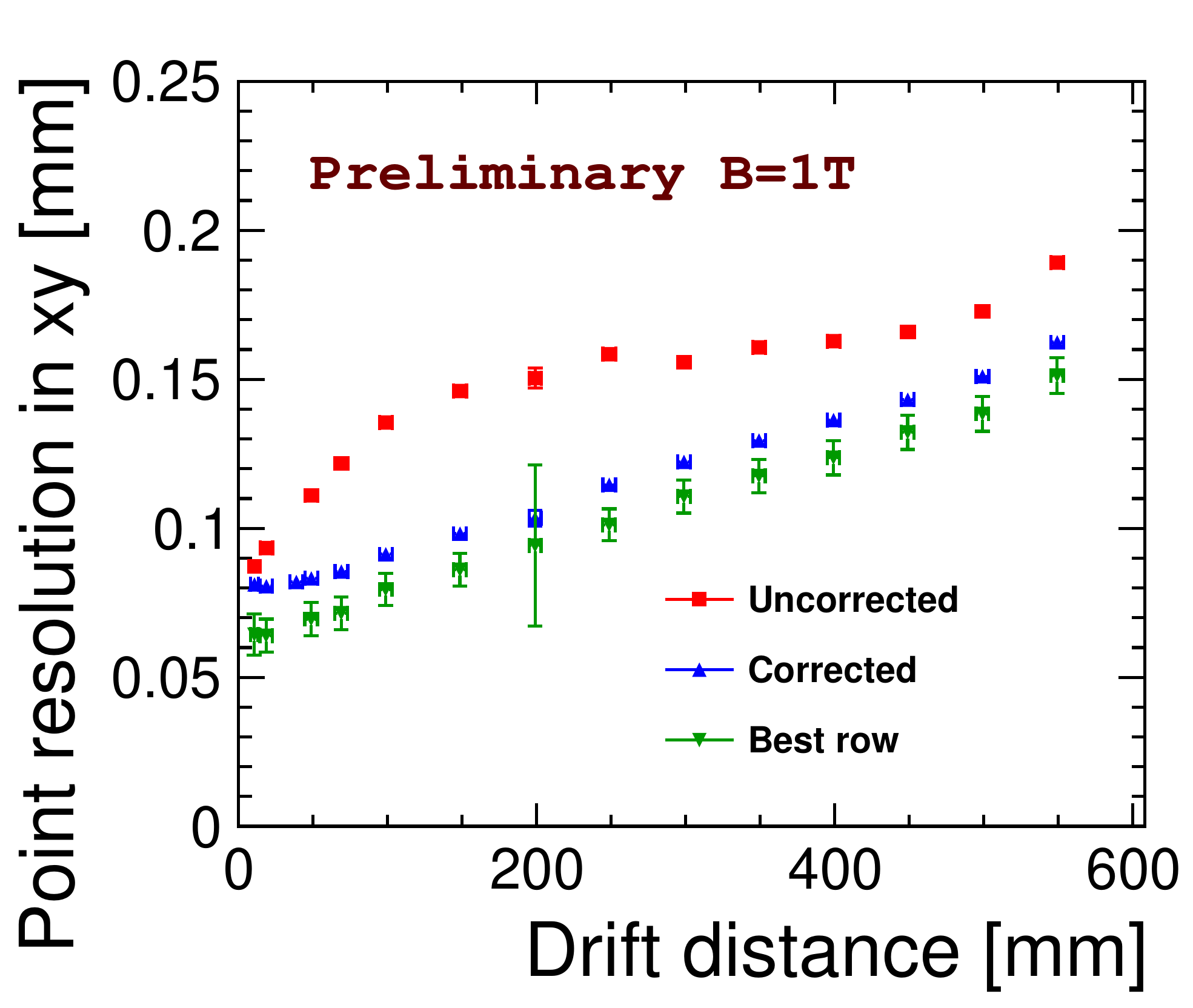}
\caption{Point resolution without magnetic field (left) and with 1T (right) with and without attempting to correct the field distortion. In addition the resolution of a single row is shown, in this case row 16 on the middle module.}
\label{fig:res}
\end{figure}

The result for the obtained single point resolution shown in Fig.~\ref{fig:res} are very preliminary with simple track fits and minimal cuts applied. All reconstructed hits are used. The uncorrected curve shows the full effect of the distortions on the resolution. 
The correction improves the overall values of the obtained resolution. In the case of a magnetic field of $\unit[1]{T}$ where the distortions are significantly larger than at $\unit[0]{T}$ also the shape of the resolution curve is affected by the distortions and does not follow the sqrt law in dependence on the drift distance any more. The corrected resolution curve shows an improved shape but still deviates from reaching an asymptotic sqrt law.
In order to ascertain the quality of the correction and get an approximation of a possible performance of a very good module one can also look at the resolution of a single row in a region of least distortions. From Fig.~\ref{fig:res_row} we pick row 16 on the middle module as the best performing row and evaluate its resolution which is shown as well in Fig.~\ref{fig:res}. The resolution does improve again with respect to the corrected curve, but in the case of $\unit[1]{T}$ it still does not lead to the expected asymptotic behavior. Possible reasons for that are that the diffusion occurring over the available drift distance is not large enough yet with respect to the pad width in order to reach the asymptotic part of the resolution curve. Other effects might be linked to the tracking and the method of extracting the resolution itself without an external reference track. A comparison of different tracking algorithms and extraction methods for the resolution that do not depend on a track fit are currently under study.

The track quality is influenced by the single point resolution and therefore by the field distortions. It is obvious that it is crucial to limit field distortion already in the design of the readout modules. Once the distortions and their systematics are better understood a description and/or parametrization should be obtained that can be used to correct the data in a more unbiased procedure.

\subsection*{Momentum Resolution}
\begin{figure}[t]
\centering
\includegraphics[width=0.7\textwidth]{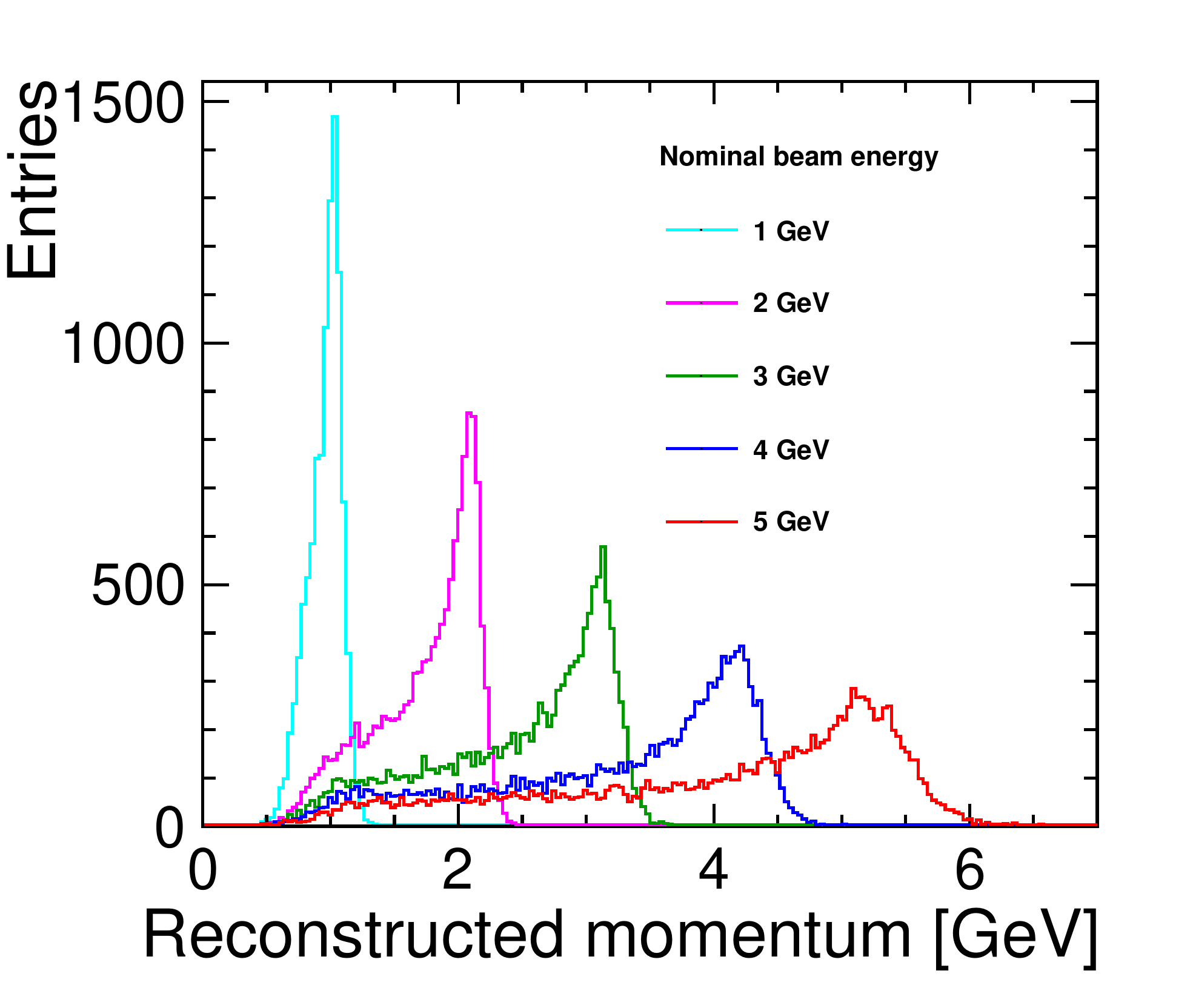}
\caption{Measured momentum distribution at different nominal beam energies. The distribution is dominated by energy loss in the magnet and beam spread.}
\label{fig:mom}
\end{figure}

Another important goal is to measure the momentum resolution.
The measured momentum distribution is dominated by energy loss in the magnet and the large beam spread as shown in Fig.~\ref{fig:mom}.
In order to obtain a momentum resolution an external tracker is required which can provide an unbiased reference track with sufficient accuracy.

In general this study and the understanding of distortions would also strongly benefit from an external reference track. Ideal would be two points before and after the TPC field cage but inside the magnet. The limited space between the TPC field cage and the inner wall of the magnet is a challenge. So far no silicon sensors with a readout system are available to fulfill those requirements.


\section{Summary and Outlook}\label{sec:sum}
A successful test beam campaign with three improved GEM modules was carried out yielding a large amount of data. The reconstruction is still being developed and improvements are made constantly. The main objective is to gain better understanding of the observed distortions and their dependencies. Work is ongoing on calibration and correction procedures. The obtained point resolution is already very close to the final goal of $\unit[100]{\micro m}$.

In order to measure the momentum resolution with sufficient precision an external track measurement is needed. The search for a suitable silicon sensor system is ongoing.

\acknowledgments
We would like to thank the LCTPC collaboration and especially the members from Lund University for
their invaluable help and support during the test beam campaign.

\end{document}